# Multimodal Modular Chain of Thoughts in Energy Performance Certificate Assessment

**Zhen Peng[1], Peter J. Bentley[2*]**
1. Department of Physics and Astronomy, University College London, Gower Street, London WC1E 6BT, UK
2. Department of Computer Science, University College London, Gower Street, London WC1E 6BT, UK

[*]**Corresponding Author:** Peter J. Bentley (p.bentley@cs.ucl.ac.uk)

Abstract:

Accurate evaluation of building energy performance remains challenging in regions where scalable Energy Performance Certificate (EPC) assessments are unavailable. This paper presents a cost-efficient framework that leverages Vision–Language models for automated EPC pre-assessment from limited visual information. The proposed Multimodal Modular Chain of Thoughts (MMCoT) architecture decomposes EPC estimation into intermediate reasoning stages and explicitly propagates inferred attributes across tasks using structured prompting. Experiments on a multimodal dataset of 81 residential properties in the United Kingdom show that MMCoT achieves statistically significant improvements over instruction-only prompting for EPC estimation. Analysis based on accuracy, recall, mean absolute error, and confusion matrices indicate that the proposed approach captures the ordinal structure of EPC ratings, with most errors occurring between adjacent classes. These results suggest that modular prompt-based reasoning offers a promising direction for low-cost EPC pre-assessment in data-scarce settings.

## 1    Introduction

Sustainability features of buildings are important for policymaking and for commercial markets such as carbon trading. Modern buildings with high-efficiency heating systems, windows, and lighting systems contribute to meeting environmental policy targets and support participation in carbon markets.

But most existing buildings are not equipped with energy-efficient equipment, hindering the achievement of worldwide environmental protection goals. A lack of reliable data on these features further hinders any organized renovation and detailed policymaking. Paid expert assessors cannot assess every property in the world, especially for properties whose owners cannot afford prohibitive assessment costs or are not motivated to pay such a large bill, while residents or landlords may not have the essential knowledge to conduct such a survey independently. The situation is even more challenging in contexts where reliable data is too sparse not only to support data-driven methods, but also to establish the local standardized assumptions required by regulatory EPC assessments (Sayfikar et al., 2025).

Recent attempts to apply Vision–Language models (VLMs) to building energy assessment are motivated by their flexibility and strong zero-shot or few-shot capabilities. However, EPC assessment is inherently a multi-attribute and hierarchical task, in which high-level ratings depend on multiple interrelated property characteristics rather than on a single visual cue. Under standard instruction-based prompting, VLMs tend to treat these attributes independently, which prevents explicit joint reasoning over multiple attributes during the inference process. For example, limited accuracy on multiple energy classification tasks under instruction-based prompts prevents practical application, although performance on very small label sets (fewer than three classes) is good (Pan et al., 2025). In this paper, this challenge is addressed by introducing a systematic, modular Chain of Thoughts reasoning framework organized as a staged inference process over interrelated building attributes.

We present a modular and multimodal Chain of Thoughts framework based on staged intermediate attribute inference. It decomposes Energy Performance Certificate (EPC) assessment into staged sub-tasks (building age, window characteristics, heating, lighting, and the aggregate EPC rating), propagates information across stages, and supplies multimodal exemplars for key tasks. This design delivers empirically validated improvements over representative baselines, as demonstrated through statistical analysis, without requiring additional model training. The framework is evaluated on publicly available records for residential properties in the United Kingdom.

The goal of this work is not to replace official EPC assessments conducted by certified experts, but to provide a low-cost, early-stage decision-support tool for preliminary screening, awareness raising, and large-scale analysis in data-sparse contexts. Accordingly, the evaluation focuses on few-shot and data-scarce conditions, reflecting realistic scenarios where large-scale labeled EPC datasets are unavailable.

The remainder of this paper is organized as follows. Section 2 reviews relevant background and related work. Section 3 describes the proposed MMCoT architecture. Section 4 details the experimental setup, and Section 5 presents the empirical results. Section 6 discusses implications and limitations, followed by conclusions in Section 7.

## 2 Background

### 2.1. Energy Performance Certificate

With growing concern over climate change, many countries and regions have enacted policies to systematically collect building energy data (Brown et al., 2002; Zou, 2019; Arcipowska et al., 2014). This process produces Energy Performance Certificates (EPCs) or equivalent energy ratings that inform real estate markets and policymaking (Office for National Statistics, 2024; Stromback et al., 2021).

EPCs play a dual role in real estate markets and public policy (Li et al., 2019; Office for National Statistics, 2024). Empirical evidence consistently demonstrates a positive association between building energy performance and rental prices across multiple national contexts, with EPCs acting as a key channel through which this positive effect is transmitted (Cajias et al., 2013; Chegut et al., 2014; Hyland et al., 2013; Stanley et al., 2016). Beyond market outcomes, EPCs also inform policymaking and regulatory guidance and are used as policy instruments to monitor progress, enable targeted retrofit programs, and guide regulatory updates (Office for National Statistics, 2024; Rijksdienst voor Ondernemend Nederland, 2024; Sveriges riksdag, 2006).

Despite these advantages, uneven capacity for EPC data collection between developing and developed regions hinders progress toward global environmental goals (Network for Greening the Financial System, 2022; Nunoo et al., 2025) and limits related cross-border financial activities (Stromback et al., 2021). The United Nations *Global Status Report for Buildings and Construction* highlights that most new buildings are being added in regions with high energy demand and limited capacity to implement energy codes, further underscoring the impact of this uneven capacity (Global Alliance for Buildings and Construction & United Nations Environment Programme, 2023). According to a Climate Bonds report, the lack of reliable EPCs in many parts of the world has already hindered the establishment of low-carbon emissions trajectories, the assessment of market performance, and the certification of assets in these regions, resulting in losses across global markets (Climate Bonds Initiative, 2023). These challenges underscore the importance of enhancing EPC computation capabilities, particularly in regions with limited capacity to implement energy codes.

These limitations in the scalability, cost-efficiency, and accessibility of existing EPC assessment mechanisms motivate the need for alternative EPC assessment approaches that can operate with minimal data, reduced cost, and limited expert intervention.

## 2.2. Conventional and data-driven approaches
### 2.2.1. Regulatory and expert-based assessment

The conventional regulatory EPC assessment process is primarily conducted using simplified existing-dwelling energy assessment methodologies, which rely on on-site surveys to collect key property data rather than detailed construction plans. Such approaches have been widely implemented in many countries and regions including the UK, the EU and Australia. (Department for Communities and Local Government, 2017; European Parliament & Council of the European Union, 2024; Australian Government, 2024)

These methodologies require trained assessors to conduct on-site inspections, recording parameters such as building age, window characteristics, heating systems, and lighting, often supported by photographic evidence. The collected information is then used as input to standard calculation procedures, such as the Reduced Data Standard Assessment Procedure (RdSAP) in the UK, to derive EPC ratings under a set of predefined assumptions regarding thermal properties, construction forms, and system performance (Department for Communities and Local Government, 2017).

Despite their simplified nature, such assessment workflows typically require 30–60 minutes of on-site data collection and incur assessment fees in the range of £60–£120 per dwelling (National Careers Service, n.d.; HomeOwners Alliance, n.d.). As a result, the labor-intensive and expertise-dependent nature of regulatory EPC assessments is widely recognized as a key barrier to scalability and large-scale deployment, particularly in resource-constrained and developing regions (Osei-poku et al., 2025). In addition to these operational constraints, the lack of reliable statistical data or locally defined default parameters in these contexts often hinders the establishment of local standardized assumptions, thereby creating an additional barrier to the deployment of regulatory EPC assessments (Sayfikar et al., 2025).

### 2.2.2. Data-driven approaches

Prior studies have explored data-driven approaches to reduce the labor requirements of EPC assessments. These works have applied a range of machine learning models to estimate building energy performance from structured inputs. Such approaches typically rely on large-scale labeled datasets, which often comprise tens of thousands of buildings, and may require substantial computational resources to support model training and optimization (Olu-Ajayi et al., 2023; Galli et al., 2022; Afzal et al., 2024).

Although these studies demonstrate that machine learning can be effective under favorable conditions, such approaches typically rely on the availability of large amounts of high-quality labeled data, which limits their practical deployment in developing regions where such data are scarce or unavailable (Khalil et al., 2022). Consequently, the data collection and model training requirements of conventional machine learning approaches pose significant challenges to scalability, motivating the exploration of methods capable of operating under zero-shot or few-shot assumptions.

In summary, while both expert-based assessment and data-driven machine learning approaches have demonstrated effectiveness in data-rich contexts, neither is well suited to scenarios characterized by limited labeled data and constrained resources. This gap motivates the exploration of alternative paradigms capable of operating under zero-shot or few-shot assumptions.

### 2.3. Applications of Vision-Language Models (VLMs)

In many developing regions, the lack of systematic EPC data collection has resulted in severe data scarcity, limiting the feasibility of training conventional computer vision models or implementing regulatory EPC assessments. Under such conditions, the zero-shot and few-shot reasoning capabilities of Vision–Language models (VLMs) offer a promising alternative, as they enable semantic understanding and inference without task-specific training data (Ding et al., 2022; Zhou et al., 2022a, 2022b; Lai et al., 2025; Xu et al., 2022).

Recent studies have demonstrated the potential of VLMs for EPC-related tasks, and in some cases reported performance comparable to or exceeding manual assessments (Bentley et al., 2024; Pan et al., 2025). However, existing work also reveals clear limitations. For example, Pan et al. showed that under standard instruction prompting, VLMs exhibit substantially reduced accuracy when applied to multi-grade energy efficiency classification tasks, performing markedly worse as the number of classes increases (Pan et al., 2025). Given that most EPC schemes involve multiple efficiency grades, such limitations constrain the direct applicability of naïve prompting strategies.

Importantly, these limitations do not primarily stem from insufficient model capacity, but rather from a mismatch between single-shot prediction paradigms and the procedural nature of EPC assessment. As a complex multi-step reasoning process, EPC assessment inherently involves the identification and aggregation of multiple intermediate parameters, a characteristic shared with other tasks that benefit from structured intermediate reasoning rather than direct end-to-end prediction (Wei et al., 2022). Moreover, improving performance through additional model training or task-specific supervision is often impractical in data-scarce regions. Consequently, there is a need for approaches that restructure VLM reasoning to better align with official EPC assessment workflows.

To address this challenge, this paper proposes integrating the manual EPC survey process into VLM-based reasoning by decomposing the assessment into key parameter identification stages and aggregating intermediate results through a structured, multi-stage reasoning process.

## 3 Multimodal Modular Chain of Thoughts

### 3.1. Problem Setting and Task Formulation

In practical EPC assessment workflows, particularly at early decision-making stages and in resource-constrained contexts, detailed building information and on-site measurements are often unavailable. Stakeholders such as property owners, tenants, or local authorities may only have access to a small number of publicly available images and coarse building descriptions, yet still require indicative estimates of energy performance to support early-stage renovation and construction planning, high-level policy screening, or decision-making in regions lacking the capacity for systematic EPC data collection (Batish & Agrawal, 2019; European Investment Bank, 2022). In this study, we focus on this early-stage, data-sparse setting and formulate EPC assessment as a decision-support problem rather than a formal certification task. The goal is to infer key building attributes and an indicative EPC rating using only visual information that can be obtained without expert knowledge or specialized equipment. Accordingly, the problem is formulated as a set of intermediate visual classification tasks over key building attributes, followed by an indicative EPC rating estimation.

Following this formulation, the assessment task is abstracted into a limited number of high-level building attributes that capture the core logic of simplified energy assessments for existing dwellings. In line with official UK EPC guidance, key attributes such as building age, window characteristics, and heating systems constitute the primary factors considered in this study (Department for Communities and Local Government, 2017). Parameters that are difficult to inspect directly, such as wall insulation, are incorporated implicitly through correlated cues such as building age, following standard assessment assumptions (Department for Communities and Local Government, 2017). In addition, lighting systems are included as supplementary information, as visual cues related to lighting are commonly present in interior images and are therefore relatively easy to obtain and infer. Based on this abstraction, EPC assessment is formulated as a sequence of five estimation tasks: building age, window characteristics, heating system type, lighting efficiency, and the final EPC rating.

Building on this task formulation, Multimodal Modular Chain of Thoughts (MMCoT) introduces a staged multimodal inference framework. The framework is organized as a modular reasoning pipeline, in which intermediate attribute predictions are explicitly reused across stages via structured information propagation. Additionally, MMCoT allows for the inclusion of multimodal few-shot exemplars at specific stages to provide visual reference cues for certain high-impact intermediate attribute estimations. Figure 1 presents a system-level overview of MMCoT, illustrating the staged pipeline, cross-stage information propagation, and the selective adoption of multimodal few-shot exemplars.

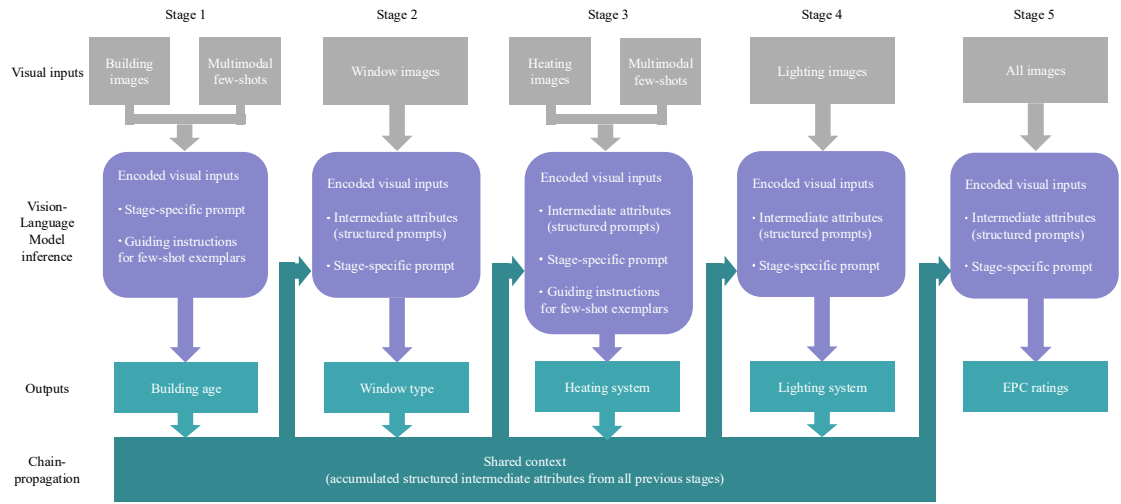

Figure 1. System-level overview of MMCoT. MMCoT operates as a structured and staged inference pipeline where property attributes are estimated sequentially from visual inputs. Intermediate attribute predictions are accumulated and propagated as structured context. Multimodal few-shot exemplars are selectively adopted in the first and third stages. Detailed descriptions of each stage and the underlying inference procedure are provided in Section 3.3.

### 3.2. Dataset and preprocessing

This study is based on a ground-truth multimodal dataset of 81 apartments with confirmed EPC records. Records were obtained from the Energy Performance of Buildings Data: England and Wales (December 2024 update) via OpenDataCommunities, and images were sourced from public listings on several rental and sales platforms. Images were manually matched to EPC records by property address. Each Energy Performance Certificate provides building age, window characteristics, main and secondary heating systems, the proportion of low-energy lighting, and related fields. The dataset is used solely for non-commercial academic research and is processed under the United Kingdom's fair-dealing provisions.

To simplify the estimation task, we define task-specific categorical labels for each intermediate attribute and for the final EPC rating. Specifically, building age is categorized into six classes: before 1900, 1900–1930, 1930–1950, 1950–1970, 1970–1990, and 1990–2020. Window characteristics are classified into three categories: single-glazed, double-glazed, and triple-glazed. Heating systems are grouped into three classes: fireplace, boiler, and electric heater. Lighting systems are categorized into five ranges according to the proportion of low-energy lighting (0–20%, 20–40%, 40–60%, 60–80%, and 80–100%). EPC ratings are classified into seven standard categories from A to G.

Although modest in size, the dataset is designed to reflect realistic EPC assessment scenarios under limited data availability. It covers the major residential property categories and building ages in the UK, with the distributions of key property types and ages (see Figure 2 (a) and (b)) broadly consistent with national housing statistics (Office for National Statistics, 2023). To facilitate robust evaluation, the dataset balance is mildly adjusted in low-frequency bands to avoid extreme class imbalance, while remaining aligned with reported EPC distribution trends (Ministry of Housing, Communities and Local Government, 2021). Given that MMCoT targets few-shot EPC prediction, the dataset is sufficient to evaluate the robustness and relative effectiveness of structured prompt-based reasoning under low-data conditions.

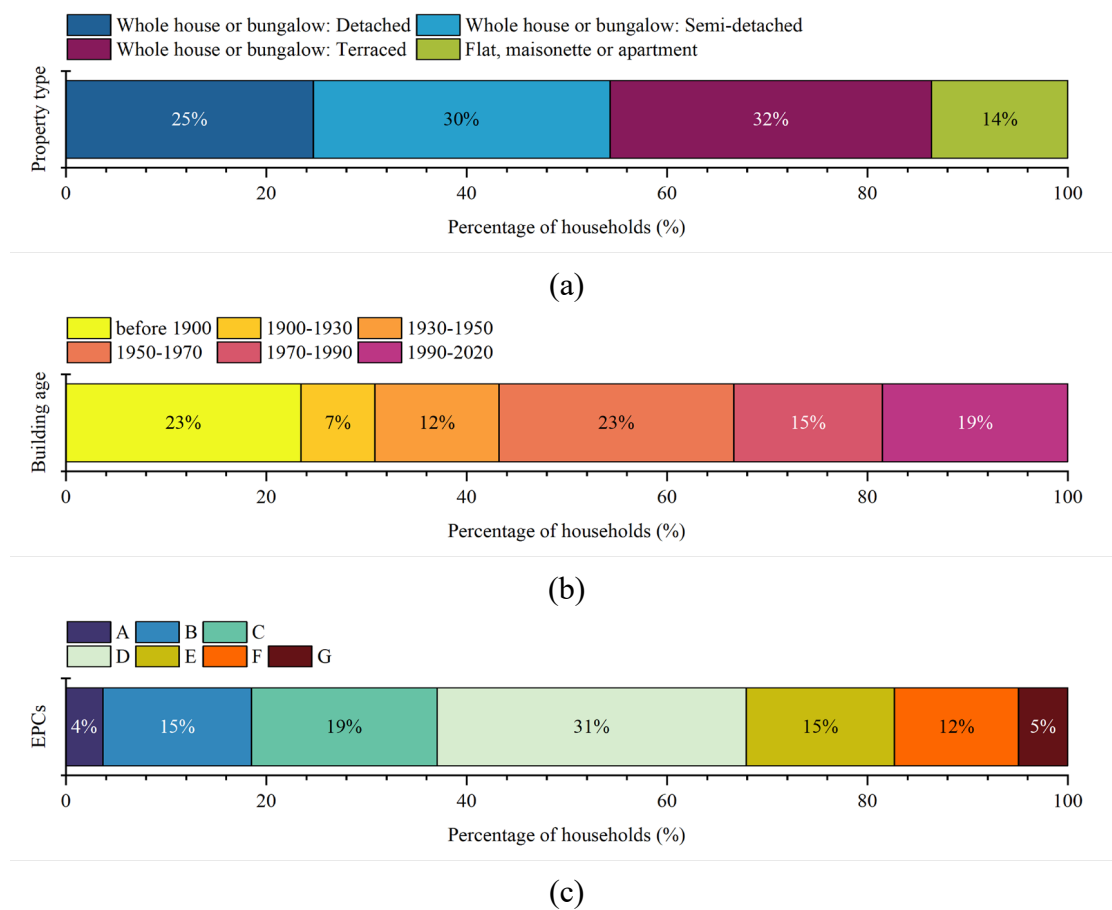

Figure 2. Property distribution in the dataset

Although the EPC records and collected images are generally reliable, basic preprocessing is applied to ensure data quality and stable experimentation. Specifically, images that are excessively blurry, heavily occluded, or dominated by irrelevant background clutter are excluded. Records lacking images of heating systems or windows are removed. Close-up views of heaters and windows are manually cropped and saved for downstream analysis, and any privacy-sensitive information is removed or masked.

### 3.3. Methodology

This section details the MMCoT inference procedure, which models property-level EPC assessment as a structured sequence of visual attribute estimation tasks, following the problem formulation introduced above.

### 3.3.1. Overview

MMCoT follows a fixed, single-pass staged inference procedure. Given a set of property images, building attributes are estimated sequentially in a predefined order, and structured intermediate predictions from earlier stages are explicitly propagated forward to subsequent stages as contextual inputs. This staged design allows later estimations to condition explicitly on previously inferred attributes, reflecting the logical dependencies commonly assumed in simplified EPC assessment workflows.

To support consistent information reuse across stages, all stages are implemented using a unified and shared prompt template (Figure 3), which provides a common execution scaffold across the pipeline. The template supports two optional functional components that are reused throughout the staged inference process: (1) a chain-propagation component, which enables intermediate attribute predictions from earlier stages to be explicitly injected as contextual inputs into subsequent stages; and (2) a multimodal few-shot component, which allows reference images to be included alongside the target instance to support visual perception in selected stages. The instantiation of this template at each stage is described in detail in the following section.

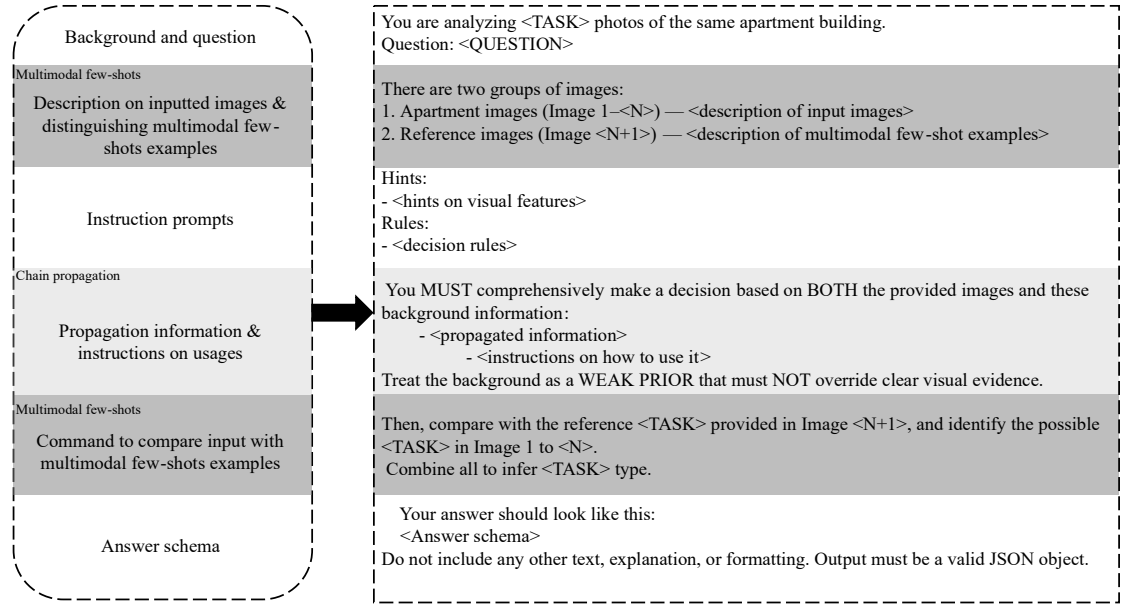

Figure 3. Prompt template for MMCoT. Light grey blocks correspond to the chain-propagation component and dark grey blocks to the multimodal few-shot component.

The modular design allows each component to be independently added or removed for ablation.

### 3.3.2. Staged Inference Procedure

For each property, MMCoT applies the staged inference framework outlined in Section 3.3.1, which defines a common structure shared across different stages. Within this framework, the system sequentially performs five attribute estimation stages: building age, window characteristics, heating system type, lighting efficiency, and final EPC rating. While the overall procedure remains consistent across instances, the stages differ in their specific inference goals and the way prompt components are instantiated. The remainder of this section describes each stage in detail, following the exact order of execution implemented in the system.

Unless otherwise specified, all stages follow the structured output format illustrated in Appendix 1 (Figure 9), and the complete stage-specific prompts are provided in Appendix 2 (Figure 10).

### Stage 1: Building Age Estimation

The first stage of the MMCoT pipeline classifies the target dwelling into one of the predefined building age categories described in Section 3.2, based solely on visual cues from property images. This stage provides a foundational attribute for subsequent estimation stages, as building age imposes strong prior constraints on other building characteristics.

For each property, only one exterior façade image from the dataset is encoded in base64 format and provided as visual input to the VLM. A stage-specific textual instruction (prompts provided in Appendix 2, Figure 10), fixed across all properties, guides the model to predict exactly one building age category from the predefined set by directing attention to architectural cues relevant to construction period estimation, such as façade materials, window proportions, roof forms, and decorative elements. While the instruction is expressed in natural language, it enforces deterministic decision constraints and requires the model to return a single categorical label in a predefined format (samples provided in Appendix 1, Figure 9). No intermediate attributes are available at this stage, as it is executed at the beginning of the pipeline.

In MMCoT, the few-shot component is designed as an optional augmentation rather than a default prompt element. By default, staged inference proceeds without additional exemplars. The component is selectively activated at stages where the underlying RdSAP procedure does not permit the parameter to be completed through standardized conventions and instead requires explicit visual determination or documentation (Department for Communities and Local Government, 2017). In MMCoT, such conventions are approximated through structured information propagation across stages. When no convention-based structural completion is available, inference depends primarily on perceptual discrimination, increasing ambiguity and the risk of downstream error propagation. Under these conditions, multimodal few-shot exemplars are introduced as a perceptual anchoring mechanism rather than as a knowledge-injection strategy, preserving alignment with the institutional logic of simplified EPC assessment procedures.

Building age satisfies this activation condition because it cannot be derived through standard assumption-based conventions within the RdSAP framework. The age band serves as the indexing variable for age-dependent conventions and must therefore be explicitly determined prior to the application of age-dependent conventions. As such, it functions as a structurally critical anchor in the assessment procedure. The introduction of multimodal few-shot exemplars (see Figure 4) stabilizes early-stage prediction by reducing the likelihood of initial misclassification and mitigating downstream error propagation. These exemplars are not used for model adaptation but to align observed architectural features with representative construction periods.

At the end of this stage, the predicted attribute label is stored as a structured intermediate attribute for use in later stages.

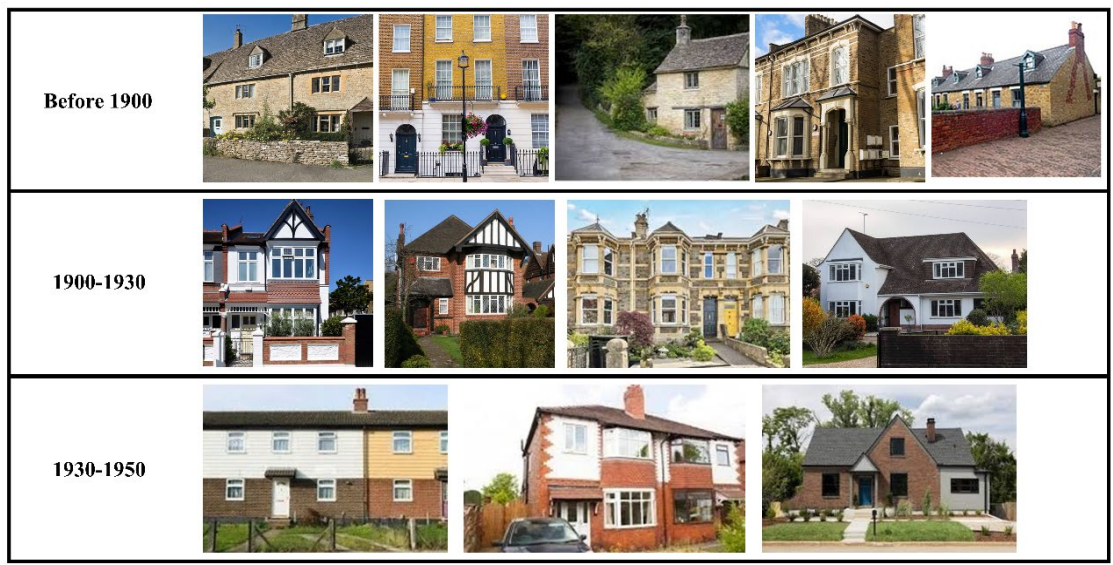

(a)

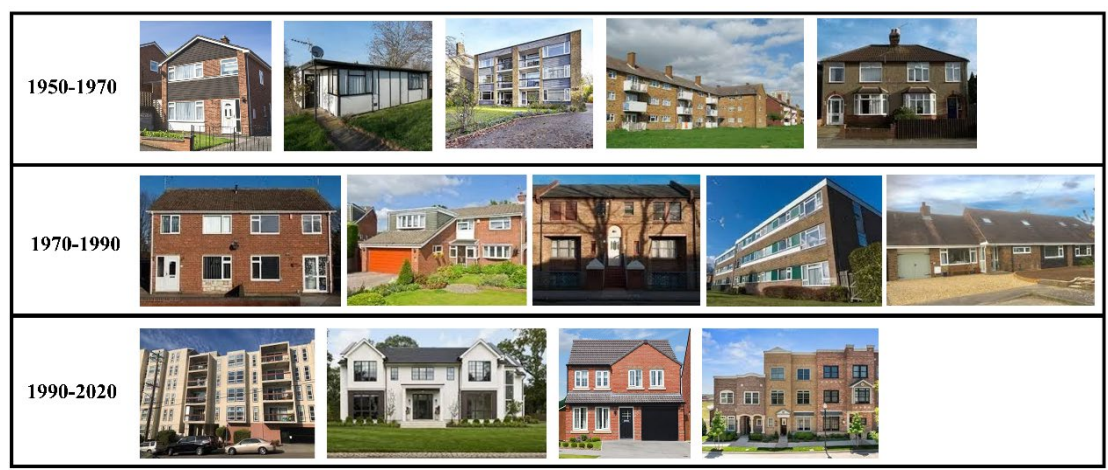

(b)

Figure 4. Multimodal few-shot exemplars for building age estimation.

**Stage 2: Window Characteristics Estimation**

The second stage of the MMCoT pipeline estimates window characteristics of the target property. This stage builds directly on the output of Stage 1 by conditioning its inference on the building age attribute inferred previously, in addition to the original property images.

For each property, only one close-up window image from the dataset is encoded in base64 format and provided as visual input to the VLM. The building age attribute predicted in Stage 1 is injected into the shared prompt template as contextual input, allowing the model to condition its estimation on age-dependent architectural conventions. A stage-specific textual instruction, fixed in structure across properties but instantiated with property-specific inferred attributes, constrains the model to predict exactly one window category from the predefined set described in Section 3.2. The instruction directs attention to visually salient window-related cues, including window framing materials, glazing patterns, sash configurations, and frame thickness, while enforcing a deterministic categorical output in a predefined format.

The prompt also includes explicit instructions on how propagated intermediate attributes should be used during inference. In particular, it encodes qualitative relationships between building age and window characteristics as contextual guidance, reflecting the general tendency for older buildings to exhibit fewer glazing layers and for newer buildings to employ more layered glazing. These relationships are provided as non-deterministic background information rather than decision rules. All propagated attributes are treated as weak priors and must not override clear visual evidence, ensuring that visually salient cues take precedence when conflicts arise and mitigating error accumulation across stages.

In contrast to Stage 1, multimodal few-shot exemplars are not introduced at this stage. Within the RdSAP framework, glazing-related characteristics may be inferred through standardized age-dependent conventions when direct evidence is unavailable (BRE Group, 2025b). As such, this parameter does not require explicit perceptual anchoring under the activation principle articulated above. The few-shot component is therefore not activated. The predicted window characteristic label is stored as a structured intermediate attribute for subsequent stages.

**Stage 3: Heating System Identification**

The third stage of the MMCoT pipeline identifies the primary heating system type of the target property, which constitutes a core determinant of overall energy performance. This stage conditions its inference on both original property images and the intermediate attributes inferred in previous stages.

For each property, one or two close-up images of heating-related components are provided as visual input to the VLM, depending on whether one or two distinct heating systems are present. The building age attribute inferred in Stage 1 and the window characteristic attribute inferred in Stage 2 are injected into the shared prompt template as contextual inputs. These propagated attributes allow the model to condition its estimation on age-dependent and envelope-related constraints that commonly co-occur with specific heating system configurations.

In addition to being provided as categorical labels, propagated intermediate attributes are accompanied by qualitative background relationships that contextualize their relevance to heating system identification. For example, building age is associated with characteristic heating system tendencies, while window characteristics provide complementary envelope-related context. These relationships are encoded in the prompt as non-deterministic guidance rather than decision rules. As specified in Stage 2, all propagated information is explicitly treated as a weak prior and must not override clear visual evidence from heating-related components.

A stage-specific textual instruction, fixed in structure across properties but instantiated with property-specific inferred attributes, provides integrated textual descriptions of diagnostic visual cues associated with different heating system configurations, including both indicative and exclusionary cues. The instructions constrain the model to infer up to two heating system categories from the predefined set described in Section 3.2 and to return the prediction in a structured format.

Multimodal few-shot exemplars are introduced at this stage (see Figure 5), as heating system identification satisfies the activation condition defined above. Within the RdSAP 10 framework, heating parameters are specified through explicit system inputs rather than age-based conventions (BRE Group, 2025a). Accurate identification of heating configuration therefore becomes structurally important. To support this requirement, few-shot visual exemplars are incorporated alongside prompt-based instruction, with representative system configurations provided as reference images.

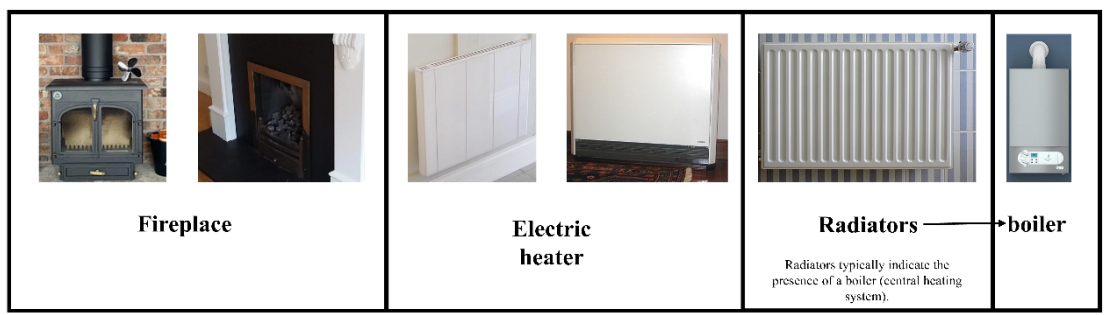

Figure 5. Multimodal few-shot exemplars for heating system identification. In the UK residential context, radiators are commonly associated with boiler-based heating systems, according to national housing statistics (Ministry of Housing, Communities and Local Government, 2021).

At the end of this stage, the predicted heating system label is stored as a structured intermediate attribute for subsequent stages.

**Stage 4: Lighting Efficiency Estimation**

The fourth stage of the MMCoT pipeline estimates the lighting efficiency of the target property. This stage conditions its inference on the original property images and the intermediate attributes inferred in previous stages.

For each property, only one interior image containing visible lighting fixtures is encoded in base64 format and provided as visual input to the VLM. The building age attribute, window characteristic attribute, and the heating system attribute inferred in the previous stages are injected into the shared prompt template as contextual inputs.

These propagated attributes provide contextual constraints that support consistent interpretation of lighting-related visual cues.

In addition to being provided as categorical labels, propagated intermediate attributes are accompanied by qualitative background constraints that contextualize their relevance to lighting efficiency estimation. In particular, building age is presented as an imperfect indicator of renovation status, with the prompt noting that energy-efficient lighting is more directly associated with recent renovations rather than building age alone. Window characteristics and heating system attributes are included as supplementary contextual information without imposing explicit directional relationships. Uncertainty handling for propagated attributes follows the same weak-prior strategy defined in Stage 2.

A stage-specific textual instruction, fixed in structure across properties but instantiated with property-specific inferred attributes, specifies textual descriptions of diagnostic visual cues relevant to lighting efficiency assessment. The instruction constrains the model to infer a single lighting efficiency category from the predefined set described in Section 3.2 and to return the prediction in a structured, machine-readable format (samples provided in Appendix 1, Figure 9).

As this stage does not meet the activation condition defined above, multimodal few-shot exemplars are not introduced. Within the RdSAP framework, lighting efficiency can be completed through predefined assumption-based procedures when precise evidence is unavailable (BRE Group, 2025a). As such, it does not require explicit perceptual anchoring under the activation principle articulated above. The predicted lighting efficiency label is stored as a structured intermediate attribute for subsequent stages.

**Stage 5: EPC Rating Inference**

At the last stage, MMCoT pipeline estimates the EPC rating of the target property. This stage conditions its inference on used images and predicted attributes in previous stages.

For this stage, all previously used images covering exterior and interior façade, window, and heating system are provided as visual input to the VLM. This design enables VLM to take all related visual cues, ensuring the completed information is inputted. The attributes on building age, window characteristics, the heating system, and the lighting system, inferred in the previous stages are injected into the shared prompt template as contextual inputs. These propagated attributes provide contextual constraints that support consistent interpretation of EPC-related visual cues.

In addition to being represented as categorical labels, the propagated intermediate attributes are accompanied by explicitly injected qualitative background constraints, which serve as weak priors to contextualize their relevance to EPC rating estimation. These constraints are not treated as deterministic rules, but as soft guidance applied only when visual evidence is insufficient or ambiguous.

Specifically, four types of qualitative background knowledge are incorporated: (1) Building age, reflecting the general tendency that newer buildings are more likely to achieve higher EPC ratings; (2) Window glazing type, reflecting a qualitative relationship whereby higher glazing levels generally contribute to improved EPC performance; (3) Heating system type, incorporated through a qualitative efficiency hierarchy that serves as a weak inductive bias for EPC estimation; and (4) Energy-efficient lighting, incorporated through an explicit positive association between the prevalence of low-energy lighting and EPC outcomes.

Uncertainty in the propagated attributes is handled using the same weak-prior strategy adopted in earlier stages, ensuring that noisy or imprecise predictions influence the final EPC categorization only in the absence of decisive visual cues.

A stage-specific instruction template, fixed in structure across properties but instantiated with property-specific inferred attributes, specifies textual descriptions of diagnostic visual cues relevant to EPC rating assessment. The instruction constrains the model to infer a single EPC rating category from the predefined set described in Section 3.2 and to return the prediction in a structured, machine-readable format (samples provided in Appendix 1, Figure 9).

No multimodal few-shot exemplars are introduced at this stage. Although the EPC rating represents the final output of the pipeline, it functions as an aggregation stage that integrates previously inferred attributes rather than serving as an independent inference anchor within the RdSAP procedure. As this stage does not introduce new parameters requiring explicit determination, it does not satisfy the activation condition defined above. The EPC category is therefore inferred using the standard staged prompt configuration.

## 4 Experiments

The following experiments build directly on the methodology described in Section 3 and evaluate MMCoT under consistent, data-sparse EPC assessment conditions.

### 4.1. Experimental Scope

In this paper, we evaluate the proposed method primarily against representative zero-shot and few-shot approaches, ensuring a fair comparison under a consistent low-data setting. The goal of this study is to develop an EPC approximation method using only limited, easily obtainable information for decision support in regions where historical labeled data are scarce or entirely unavailable.

Traditional supervised machine learning models fundamentally rely on substantial amounts of labeled training data and are therefore outside the methodological scope of this work. Likewise, conventional rule-based EPC computation typically requires detailed property information obtained through professional surveys and expert diagnosis, which is incompatible with the lightweight assessment setting considered here. Therefore, fully supervised EPC predictors and standard rule-based EPC pipelines are not included as direct baselines in this paper.

To facilitate meaningful comparison under identical input constraints, we additionally consider constrained proxy baselines that approximate rule-based or learning-based approaches while operating on the same limited information available to MMCoT. In particular, we include a constrained baseline modified from RdSAP, the standardized methodology underlying official EPC assessments in the UK, operating on the same intermediate predicted attributes as MMCoT, with remaining missing inputs completed using standardized assumptions. In addition, we include a lightweight few-shot baseline based on linear probing over frozen VLM representations, which provides an ML-style reference without requiring end-to-end supervised training. Further implementation details are described in Section 4.2.

In addition to baseline comparisons, we conduct ablation and control experiments to isolate and assess the contribution of individual components in MMCoT.

### 4.2. Baselines

The proposed method is compared against four baselines spanning rule-based approximation, zero-shot Vision–Language models, and lightweight few-shot classifiers. The implementations of baselines are listed below:

(1) EPCTK-based RdSAP Baseline. This baseline is established using the open-source computation tool EPCTK, which follows the RdSAP methodology underlying official EPC assessments in the UK (Chambers, J. D., 2017). To ensure identical input availability, this baseline operates on the same set of predicted intermediate attributes as MMCoT, without inheriting its final EPC inference or staged reasoning process. Remaining variables are completed using standardized RdSAP defaults or national average values from UK statistics.

(2) Voting-based zero-shot CLIP. This baseline leverages CLIP, a Vision-Language model operating in a shared multimodal embedding space. Given multiple images associated with corresponding instructions, CLIP performs EPC reasoning independently for each input. The final EPC level is obtained by aggregating these individual predictions via a voting strategy.

(3) CLIP with Logistic Regression. In this baseline, multiple images are aggregated to generate an embedding by CLIP. A logistic regression classifier implemented in scikit-learn is trained on top of these embeddings with a maximum of 1000 iterations and default L2 regularization. The classifier is evaluated using 3-fold cross-validation. Given the limited dataset size, this classifier is not intended to represent a fully supervised solution, nor to compete with traditional supervised EPC predictors, but rather serves as a lightweight few-shot reference built on frozen CLIP representations.

(4) Zero-shot GPT. In this baseline, the same images, instruction prompts, and hyperparameter settings as MMCoT are used. However, this baseline does not include multimodal few-shot exemplars and the chain propagation component. It therefore serves as a zero-shot reference using the same VLM.

### 4.3. Ablation and Control Experiments

To assess the effectiveness of components in the proposed method, we further conduct an ablation study through property-level controlled experiments, in which model performance is evaluated across all tasks. This setting captures how variations in task-level predictions relate to the effectiveness of the overall reasoning process, rather than being restricted to the final EPC estimation. For each experimental run, the prompt template (except for the key components described in Section 3.3), hyperparameters, ground truth, provided images, apartment order, model, and runtime environment are held constant. Four experimental settings are implemented: (1) a baseline using instruction prompts; (2) instruction prompts augmented only with chain propagation; (3) instruction prompts augmented only with multimodal few-shot exemplars; and (4) the full Multimodal Modular Chain of Thoughts (MMCoT). All other prompt fields were identical across settings, and only the specified components were toggled. Figure 3 illustrates the structured prompt template of MMCoT and shows how it is possible to add and remove each component independently for ablation. These controlled settings form a rigorous ablation framework, providing evidence for the effectiveness of the different components in MMCoT.

Beyond the prompt-control experiments, a propagation-control experiment was conducted to approximate MMCoT performance when visual perception fails and to test the importance of propagation. In this control, all propagated answers were replaced with random strings of the same length as the originals. If performance with random propagated answers is significantly worse than performance with predicted answers, propagation can be considered to contribute materially. Otherwise, if performance with random propagated answers is similar to or better than that with predicted answers, propagation does not contribute to performance.

In these control experiments, we propose an overall indicator to comprehensively evaluate the performance of the entire reasoning process. Specifically, for each sample $i$, we compute a macro accuracy defined as

$$Accuracy_i = \frac{1}{K} \sum_{k=1}^{K} [\hat{y}_{i,k} = y_{i,k}] \qquad (1)$$

Where $K$ denotes the number of evaluated fields and $[\cdot]$ equals 1 if the condition holds and 0 otherwise. The overall performance difference between two methods is then evaluated by comparing the set of $\{Accuracy_i\}$ across all samples.

**4.4. Implementation Details and Metrics**

In this paper, we use gpt-4o-2024-08-06 as our Vison-language Model for all experiments. The decoding temperature of the Vision-Language model is fixed at 0 to reduce stochasticity and improve reproducibility. Implementations are executed in Python using Jupyter Notebook. All predictions are evaluated against ground truth data obtained from an actual EPC assessment database, using overall classification accuracy and macro-averaged recall, as defined by the formulas below.

$$Accuracy = \frac{N_{correctness}}{N} \qquad (2)$$

$$Recall_c = \frac{TP_c}{TP_c + FN_c} \qquad (3)$$

$$Recall_{macro} = \frac{1}{N_{class}} \sum_{c=1}^{N_{class}} Recall_c \qquad (4)$$

Where $N_{correctness}$ denotes the number of correctly predicted samples, N represents the total number of samples, $TP_c$ is the number of samples correctly predicted as class c, $FN_c$ is the number of samples that belong to class c but are incorrectly predicted, and $N_{class}$ denotes the total number of classes.

## 5 Results

### 5.1. Performance

To evaluate performance of proposed method and baselines, all predictions of EPC rate were compared against the ground truth, and accuracy and recall were computed through formulas in Section 4.4. Accuracy was used to reflect the reliability of outputs from the Vision–Language model, whereas recall was used to measure the proportion of true cases that were correctly identified rather than missed. Using these two metrics, the proposed architecture was assessed.

As shown in Figure 6, MMCoT consistently outperformed all baselines across both Recall and Accuracy. Compared with the rule-based RdSAP-like method, MMCoT yields substantially higher accuracy under the same input constraints. In addition, MMCoT maintains a clear performance margin over CLIP-based zero-shot and linear-probing variants. Notably, improvements in accuracy are accompanied by consistent gains in recall, indicating that the performance boost does not come at the cost of reduced coverage. Nevertheless, the overall accuracy remains relatively low, reflecting the intrinsic difficulty of the task and leaving room for further improvements.

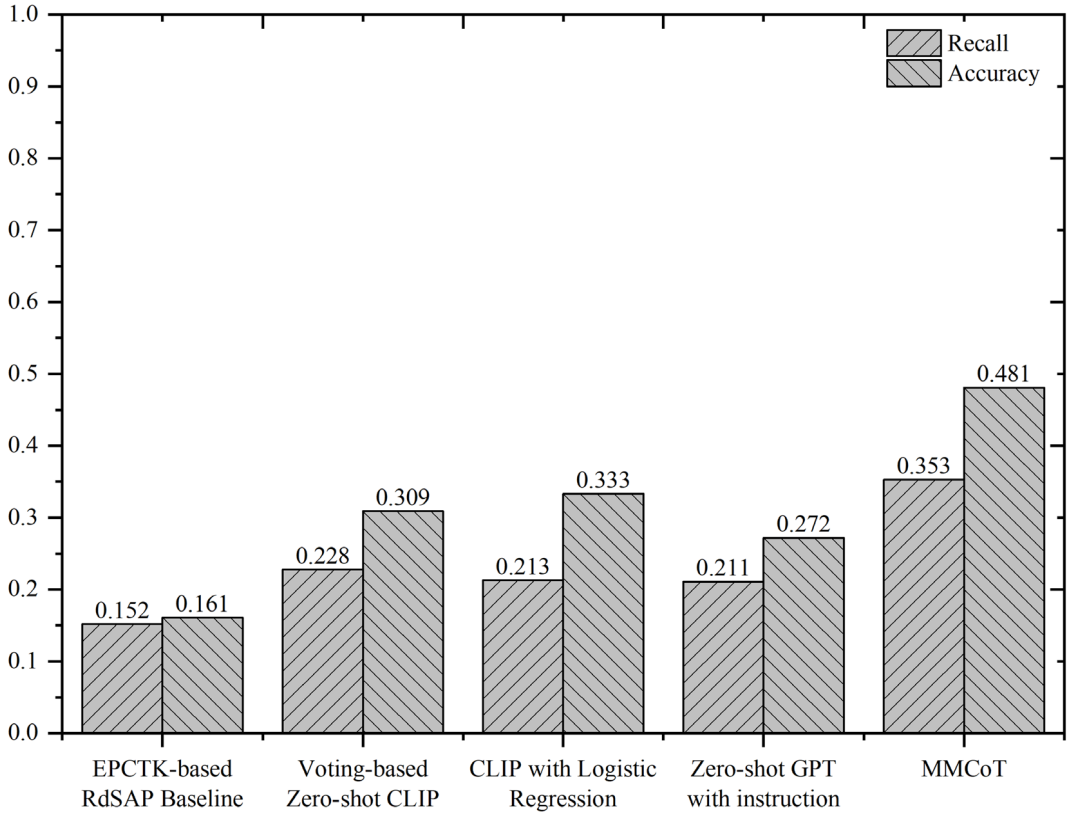

Figure 6. Comparison of Accuracy and Recall across different methodologies on EPC rate prediction.

To further evaluate prediction quality beyond classification performance, we map the predicted EPC values into a discrete sequence with a unit interval and compute the mean absolute error (MAE), which is reported in Figure 7. Generally, MMCoT achieves the lowest MAE among all compared methods. Compared with the EPCTK-based baseline, MMCoT reduces the MAE from 1.988 to 0.741, corresponding to a substantial reduction in prediction error. Even compared with CLIP- and GPT-based zero-shot baselines, MMCoT consistently yields substantially smaller prediction errors. This trend is consistent with the improvements observed in Accuracy and Recall, suggesting that MMCoT not only makes more correct predictions, but also produces more precise outputs when errors occur.

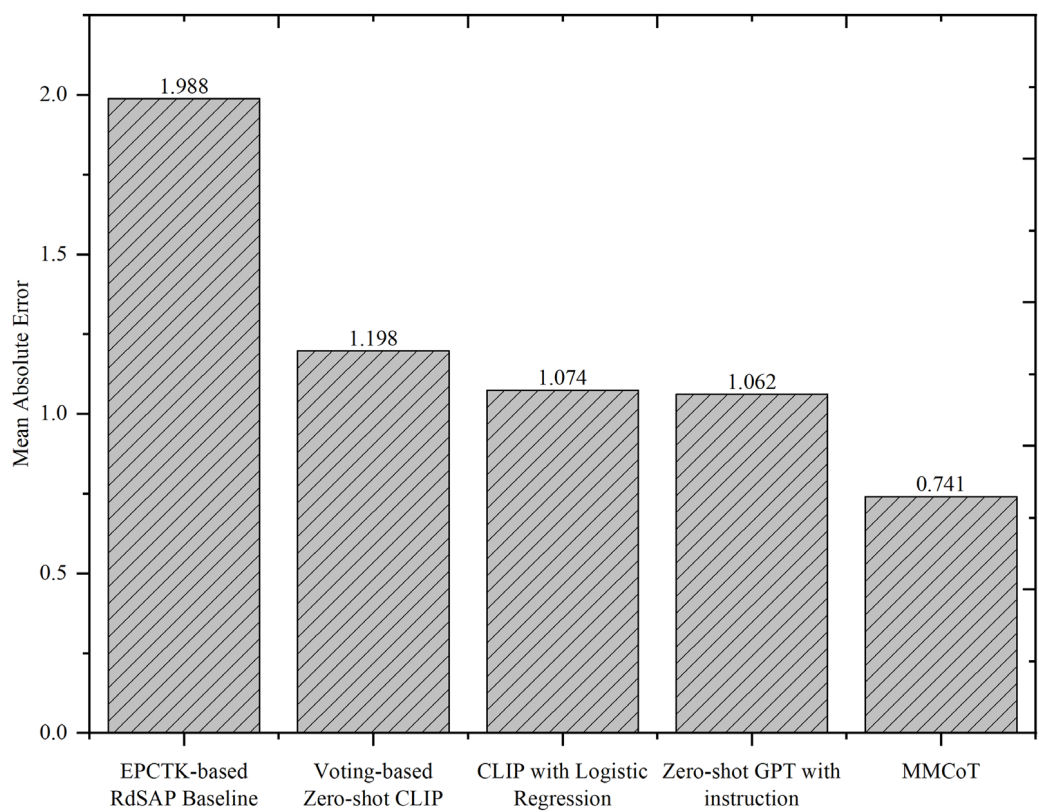

Figure 7. Comparison of mean absolute error (MAE) across different methodologies on EPC rate prediction.

The confusion matrix of MMCoT (see Figure 8) demonstrates an overall consistent classification performance, with prediction results predominantly distributed along the main diagonal. In particular, MMCoT achieves relatively high true positive rates for the middle-range classes (e.g., C and D). Most misclassifications occur between adjacent classes rather than across distant categories, indicating that prediction errors are generally limited in magnitude. However, for classes corresponding to extreme EPC rates (e.g., A and G), the classification performance is comparatively lower.

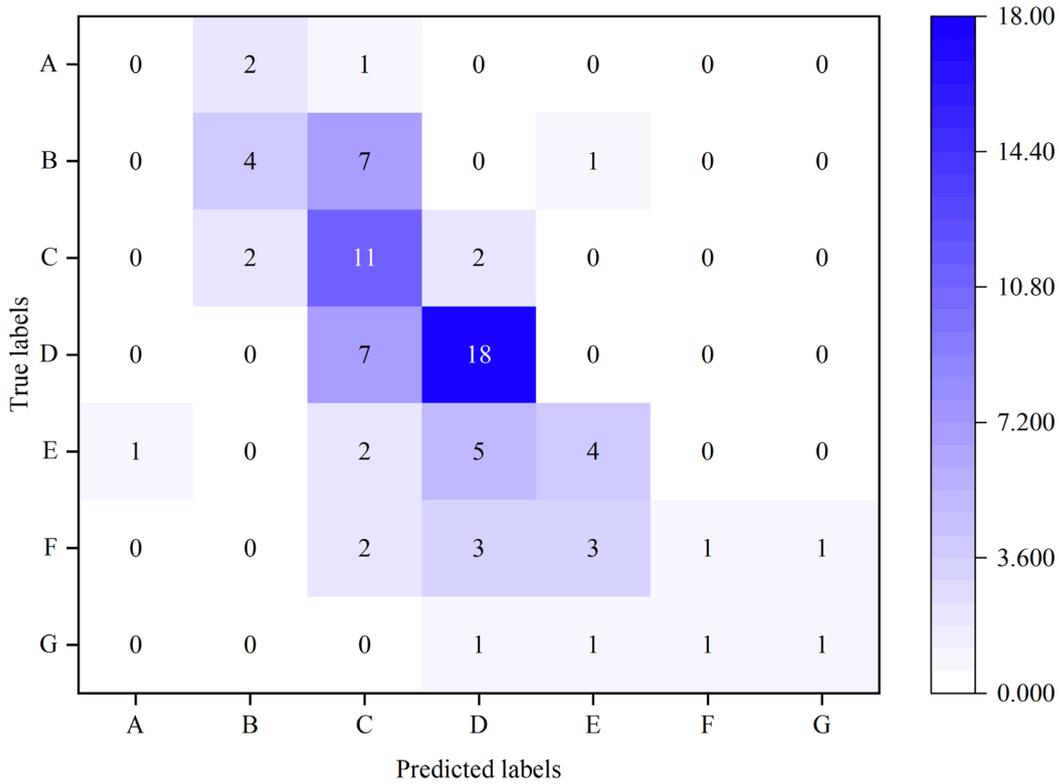

Figure 8. Confusion matrix of MMCoT

### 5.2. Ablation results

We conduct ablation experiments using statistical significance analysis via a paired bootstrap, following the indicator definitions introduced in Sections 4.3 and 4.4, with the goal of isolating the contribution of individual components in MMCoT. In addition, to better understand the effectiveness of different components in the overall reasoning process, we evaluate model predictions across all five tasks and analyze improvements in intermediate predictions. The zero-shot GPT described in Section 4.2 is selected as the baseline, as it constitutes the core component of MMCoT.

The procedure was applied to apartment-level accuracy results for MMCoT, the chain-propagation component, and the multimodal few-shot component. Relative improvements over the baseline were computed as per-property deltas, and the paired bootstrap was carried out on these deltas with 10,000 resamples under a fixed global seed. Detailed 95% confidence intervals and p-values for comparisons of each method against the baseline across all five tasks are reported in Table 1.

Compared with the baseline, MMCoT does not show statistically significant improvements on most individual sub-tasks but achieves significant gains on EPC estimation and in overall performance, providing evidence for the effectiveness of the proposed architecture. For age estimation, although the improvement is not statistically significant, the accuracy difference and confidence interval trend toward positive values. In contrast, the performance differences on heating and lighting system estimation are negligible, and the small negative difference observed for window estimation also does not reach statistical significance.

For the chain-propagation setting, statistically significant differences are observed on EPC estimation and window estimation, with a positive gain on the former and a negative difference on the latter. Performance differences on the remaining sub-tasks as well as the overall metric are small and do not reach statistical significance.

For the multimodal few-shot setting, no statistically significant improvements are observed across any tasks. Although the differences do not reach statistical significance, age estimation and the overall metric show small positive differences, with confidence intervals overlapping zero.

**Table 1. Significance analysis across settings compared with the instruction-only baseline.**

| Model comparison | Tasks | Accuracy difference | p-value | Confidence intervals | Statistically significant |
|---|---|---|---|---|---|
| MMCoT vs. baseline | Age estimation | 0.06 | >0.05 | [-0.01, 0.14] | No |
| | Window estimation | -0.04 | >0.05 | [-0.04, 0.00] | No |
| | Heating system estimation | 0.00 | >0.05 | [0.00, 0.00] | No |
| | Lighting system estimation | 0.00 | >0.05 | [-0.07, 0.07] | No |
| | EPC estimation | 0.21 | <0.05 | [0.11, 0.32] | Yes |
| | Overall | 0.05 | <0.05 | [0.03, 0.08] | Yes |
| Chain-propagation vs. baseline | Age estimation | 0.01 | >0.05 | [-0.03, 0.05] | No |
| | Window estimation | -0.05 | <0.05 | [-0.10, -0.01] | Yes |
| | Heating system estimation | 0.00 | >0.05 | [0.00, 0.00] | No |
| | Lighting system estimation | 0.00 | >0.05 | [-0.06, 0.07] | No |
| | EPC estimation | 0.16 | <0.05 | [0.05, 0.27] | Yes |
| | Overall | 0.02 | >0.05 | [0.00, 0.05] | No |
| Multimodal few-shots vs. baseline | Age estimation | 0.07 | >0.05 | [0.00, 0.16] | No |
| | Window estimation | -0.01 | >0.05 | [-0.04, 0.00] | No |
| | Heating system estimation | 0.00 | >0.05 | [0.00, 0.00] | No |
| | Lighting system estimation | 0.02 | >0.05 | [-0.03, 0.07] | No |
| | EPC estimation | -0.01 | >0.05 | [-0.04, 0.00] | No |
| | Overall | 0.01 | >0.05 | [-0.01, 0.04] | No |

### 5.3. Propagation-control experiment

Under the same significance procedure, results were analyzed for random propagation and for propagation using predicted answers. Detailed 95% confidence intervals and p-values for all five tasks and the overall metric are reported in Table 2.

Compared with random propagation, the standard MMCoT achieves statistically significant improvements on EPC estimation and overall performance. Although the remaining differences do not reach statistical significance, lighting estimation shows a small positive difference, while age estimation and window estimation exhibit small negative differences, with confidence intervals overlapping zero.

Table 2. Significance analysis for propagation control.

| Task | Accuracy difference | p-value | Confidence intervals | Statistically significant |
|---|---|---|---|---|
| Age estimation | -0.04 | >0.05 | [-0.09, 0.01] | No |
| Window estimation | -0.01 | >0.05 | [-0.04, 0.00] | No |
| Heating system estimation | 0.000 | >0.05 | [0.00, 0.00] | No |
| Lighting system estimation | 0.025 | >0.05 | [-0.03, 0.07] | No |
| EPC estimation | 0.27 | <0.05 | [0.16, 0.38] | Yes |
| Overall | 0.05 | <0.05 | [0.02, 0.07] | Yes |

# 6 Discussion

The observed improvements in both accuracy and recall over all baselines indicate that the proposed MMCoT architecture is effective in extracting and integrating heterogeneous information under data-sparse conditions. Unlike conventional EPC prediction methods that rely on structured inputs, MMCoT operates directly on limited and unstructured visual evidence, which is often the only information available in early-stage or low-cost assessments. The relatively limited performance of the EPCTK-based RdSAP baseline largely reflects the structural dependence of conventional EPC assessment approaches on rich input data, and their limited adaptability to data-sparse scenarios. The comparable performance of zero-shot CLIP and zero-shot GPT highlights the inherent limitations of single-stage, instruction-based reasoning in this setting, while the marginal gains obtained by training a lightweight logistic regression classifier on frozen CLIP embeddings suggest that few-shot supervised adaptation alone is insufficient to address the complexity of EPC estimation. In contrast, MMCoT consistently achieves the best performance among all baselines when only basic visual data are available, underscoring the effectiveness of modular reasoning and cross-task information propagation in scenarios where traditional methods struggle due to missing or uncertain inputs.

The low mean absolute error and the confusion matrix analysis further indicate that MMCoT generally captures the ordinal structure of EPC classification. Correct predictions are predominantly concentrated in the mid-range EPC classes, while most misclassifications occur between adjacent categories rather than across distant levels, suggesting that prediction errors are typically limited in magnitude. In contrast, classification performance at the extreme EPC bands, such as A and G, is comparatively lower. This behavior can be attributed to the intrinsic uncertainty associated with these extremes, which are relatively rare, exhibit weaker visual discriminability, and are therefore more difficult to distinguish reliably based solely on property images. Notably, official UK energy efficiency analyses often aggregate extreme EPC bands, such as A/B or F/G, for statistical reporting and policy evaluation, reflecting similar challenges in achieving stable fine-grained distinctions at the boundaries. (Ministry of Housing, Communities and Local Government, 2021) From this perspective, the remaining

MMCoT errors at the extremes, which are largely confined to neighboring classes, do not indicate a fundamental misalignment with EPC assessment practice, but rather reflect the inherent instability of fine-grained classification in these regions. Such errors may therefore be considered acceptable within the intended scope of early-stage or low-cost EPC pre-assessment.

The ablation results indicate that the statistically significant improvement in EPC estimation under MMCoT arises from the combined contributions of its modular components rather than from uniform gains across all sub-tasks. Chain propagation yields statistically significant improvements in EPC estimation, which suggests that explicitly propagating intermediate predictions is particularly beneficial for tasks requiring composite reasoning. At the same time, its impact on individual attribute estimation is mixed, with negligible effects on several sub-tasks and a negative effect on window estimation, indicating that propagated uncertainty may occasionally amplify errors in visually ambiguous attributes. In contrast, the multimodal few-shot component does not produce statistically significant improvements in most tasks. However, it consistently shows positive trends in age estimation, which, although not significant in isolation, may still contribute indirectly to improved EPC prediction when combined with downstream reasoning. Taken together, these results suggest that the two components play complementary roles, with chain propagation enhancing cross-attribute reasoning and multimodal few-shot prompting improving specific intermediate attributes. Their combination leads to the overall performance gains observed for MMCoT, despite the presence of task-specific limitations.

Propagation control analysis further supported the role of information propagation. When random or misleading propagated entries were replaced with predicted answers from earlier steps, significant performance improvements were obtained. These results indicate that propagated information materially contributes to performance and that, during comprehensive reasoning, the Vision-Language model uses propagated cues to inform its decisions. The guidance introduced in the information chain propagation components did not contribute to the performance improvements independently.

As a decision-support tool designed for early-stage or data-sparse scenarios, MMCoT can be deployed at low computational and monetary cost. In our experiments, the total inference cost was under 4 US dollars for assessing over 80 properties, corresponding to approximately five cents per property. These figures are provided for illustrative purposes only, as actual costs may vary with model choice and pricing policies. In contrast, conventional EPC assessments typically require on-site expert surveys and manual data collection, which incur substantially higher costs and logistical effort. From this perspective, MMCoT offers a scalable and cost-efficient complementary solution for preliminary screening or large-scale analysis.

Although the proposed architecture achieved significant improvements on EPC-related tasks, several limitations remain. First, experiments were conducted on a dataset containing only 81 properties, which may leave rare conditions underrepresented. A larger and more diverse dataset will be required to more comprehensively evaluate the generalization potential of Vision-Language models across different conditions. Second, this study deliberately relied on easily accessible visual inputs. However, the information content of such inputs is inherently limited for fine-grained EPC assessment, particularly at the extreme efficiency bands. Future work should therefore explore additional data sources that remain easy to obtain but provide richer and more informative cues, in order to better capture subtle distinctions relevant to EPC classification. Finally, although the proposed method is motivated by deployment in data-scarce regions, quantitative evaluation necessarily requires reliable ground-truth EPC records. At the current stage, such validated benchmarks are largely unavailable in many target regions. We therefore conduct experiments in the United Kingdom, where high-quality EPC data exist, to establish the internal validity and robustness of the proposed approach under controlled conditions. Cross-regional deployment and evaluation are left for future work.

## 7  Conclusion

Although systematic collection of building design and construction data has supported the implementation of Energy Performance Certificate (EPC) policies, the high cost and limited scalability of expert-based assessments remain significant challenges. This study demonstrates that Vision–Language models can be leveraged as a supplementary, low-cost decision-support tool for EPC pre-assessment by integrating chain propagation and multimodal few-shot prompting. The proposed Multimodal Modular Chain of Thoughts (MMCoT) framework consistently outperforms representative baselines under identical input constraints, highlighting the complementary roles of structured information propagation and targeted multimodal perception in EPC-oriented reasoning.

While the present evaluation is conducted on a limited sample of residential properties in the United Kingdom, the results indicate the potential of modular prompt-based reasoning frameworks for energy performance assessment in data-scarce settings. Future work will focus on expanding the evaluation to larger and more diverse datasets, exploring additional easily accessible data sources, and assessing the applicability of the proposed approach across different regional EPC schemes.

# Appendix 1. Sample of MMCoT input and output

Figure 9 illustrates a representative example of the multimodal input and the corresponding structured output generated by MMCoT. The typical visual inputs include:

(i) an exterior view focusing on the target building,

(ii) a close-up image of a window,

(iii) a close-up image of a heating system, and

(iv) an interior image showing visible lighting systems.

MMCoT performs five predictive tasks:

(1) Building age — classified into six periods: before 1900, 1900–1930, 1930–1950, 1950–1970, 1970–1990, and 1990–2020;

(2) Window glazing — categorized as single-glazed, double-glazed, or triple-glazed;

(3) Heating system — identified as fireplace, boiler, or electric heater;

(4) Lighting efficiency — estimated in five percentage ranges: 0–20%, 20–40%, 40–60%, 60–80%, and 80–100%;

(5) EPC rating — predicted across seven standard levels: A, B, C, D, E, F, and G.

The structured prediction format returned by the model is also shown in Figure 1.

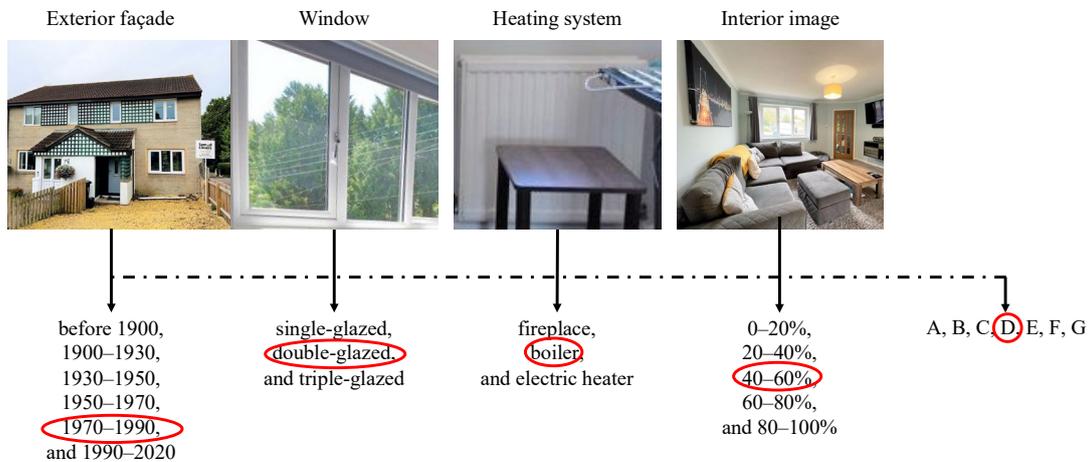

Figure 9. Sample of MMCoT input and output

# Appendix 2. Stage-specific prompts

Figure 10 presents the complete stage-specific prompts used in MMCoT across all five stages. Each prompt includes:

(i) task-specific instructions,

(ii) reasoning hints where applicable, and

(iii) explicit formatting constraints requiring the model to return predictions in a structured JSON object.

These prompts are designed to enforce structured reasoning and enable reliable chain propagation between stages.

| Stage 1 | Stage 2 | Stage 3 | Stage 4 | Stage 5 |
|---|---|---|---|---|
| You are analyzing EXTERIOR photos of the same apartment building.<br>**Question:** What is the age of this apartment?<br>**Evidence weight:** window opening > roof & chimneys > wall materials > ornament > frames<br>**Era hints:**<br>- before 1900: tall windows, arches, strong cornice, steep roofs, chimneys<br>- 1900-1930: masonry craft, bay windows, simpler decoration<br>- 1930-1950: more horizontal feel, steel-like casements, simpler roofs<br>- 1950-1970: plain blocks, low/flat roofs, repetitive small openings<br>- 1970-1990: heavier brick/concrete, ribbon-like or deeply set windows<br>- 1990-2020: neat brick/render, uniform double-glazed openings, many flat roofs<br>- 2020-now: very clean minimal joints, slim/large glazing, mixed materials<br>**Rules:**<br>- Pick ONE era best supported by most façade cues.<br>- If modern frames contradict older openings/details → choose the openings.<br>- If a tie between adjacent eras → choose the newer.<br><br>Respond **only** with a JSON object in the following format:<br>{{<br>  "age_category": "one of the following: 'before 1900', '1900-1930', '1930-1950', '1950-1970', '1970-1990', '1990-2020', '2020-now'"<br>}}<br>Do not include any other text, explanation, or formatting. Output must be a valid JSON object. | You are analyzing WINDOWS photos of the same apartment building.<br>**Question:** What is the most common window type in this apartment?<br>**Window hints:**<br>- single glazed: no spacer line; 1 reflection.<br>- double glazed: 1 spacer line; 2 reflections.<br>- triple glazed: 2 spacer lines; 3 reflections.<br>**Rules:**<br>- Use perimeter spacer-bar lines and number of reflections as primary evidence.<br>- If bar line or reflection is unclear, use apparent glass/frame depth.<br>- Tie between adjacent types → choose the moderate (double glazed).<br>You MUST comprehensively make a decision based on BOTH the provided images and these background information:<br>- The age of this apartment is: **{propagated building age}**<br>- **Single glazed**: Typically found in older buildings.<br>- **Double glazed**: More common in modern buildings.<br>- **Triple glazed**: Found in newer or energy-efficient homes.<br>Treat the background as a WEAK PRIOR that must NOT override clear visual evidence.<br>Respond only with a valid JSON object in the following format:<br>{{<br>  "window_category": "one of the following: 'single glazed', 'double glazed', 'triple glazed'"<br>}}<br>Do not include any other text, explanation, or formatting. Output must be a valid JSON object. | You are analyzing HEATERS photos of the same apartment building.<br>**Question:** Identify which of the following heating systems are visible in the images:<br>**Heater hints:**<br>- fireplace: real firebox/stove body or a visible flue/chimney.<br>- boiler: EITHER (a) wall-mounted boiler box with multiple pipes and a flue, OR (b) hydronic radiators with visible valves/pipes (counts even if the boiler unit is not shown).<br>- electric heater: wall-mounted panel heater (thin/flat, no water pipes) OR storage heater (bulky with front/top vents) OR portable convector/fan/oil-filled heater; a cord/plug/fused spur helps but is **not required**; exclude if clearly plumbed.<br>**Rules:**<br>- Mark "Y" if **any** image clearly shows features matching that category; multiple categories can be "Y".<br>You MUST comprehensively make a decision based on BOTH the provided images and these background information:<br>- The age of this apartment is: **{propagated building age}**<br>- **fireplace**: Typically found in older buildings.<br>- The window type of this apartment is: **{propagated window type}**<br>Treat the background as a WEAK PRIOR that must NOT override clear visual evidence.<br>Your answer should look like this:<br>###<br>{{<br>  "fireplace": "Y/N",<br>  "boiler": "Y/N",<br>  "electric heater": "Y/N"<br>}}<br>###<br>Do not include any other text, explanation, or formatting. Output must be a valid JSON object. | You are analyzing INTERNAL photos of the same apartment building.<br>**Question:** What type of lighting does this apartment have?<br>**Rules:**<br>- Infer the BUILDING SHELL era from photos. Prefer fixed envelope/joinery cues over replaceable items to infer building age<br>- newer building indicates higher proportion of low-energy lighting<br>Decision: If tie, weakly bias higher for clearly renovated ceilings with uniform recessed lights.<br>You MUST comprehensively make a decision based on BOTH the provided images and these background information:<br>- The age of this apartment is: **{propagated building age}**<br>- Recently renovated rooms often have energy-efficient lighting. (built age ≠ renovation age)<br>- The window type of this apartment is: **{propagated window type}**<br>- The heater system of this apartment is: **{propagated heating system}**<br>Treat the background as a WEAK PRIOR that must NOT override clear visual evidence.<br>Respond only with a valid JSON object in the following format:<br>{{<br>  "lighting_category": "one of the following: '0-20% low energy lighting', '20-40% low energy lighting', '40-60% low energy lighting', '60-80% low energy lighting', '80-100% low energy lighting'"<br>}}<br>Do not include any other text, explanation, or formatting. Output must be a valid JSON object. | You are given FOUR photo types of the SAME apartment: (1) exterior, (2) interior room, (3) heater close-up, (4) window close-up.<br>**Question:** What type of EPC does this apartment have?<br>**Rule:**<br>- **Age (weak prior):** Newer buildings usually have better EPC; use this only when other evidence is scarce or conflicting.<br>- **Windows:** Triple glazing > double > single — higher glazing levels generally improve EPC.<br>- **Heating:** boiler + hydronic radiators improve; electric heaters are moderate; fireplace lower.<br>- **Lighting:** More energy-efficient lighting improves EPC.<br>- **Provided images:** Use visible features (e.g., heater/cylinder details, glazing condition/seals, obvious vents/gaps, any PV/solar-thermal/external insulation) to support your judgment.<br>You MUST comprehensively make a decision based on BOTH the provided images and these background information:<br>- The age of this apartment is: **{propagated building age}**<br>- The window type of this apartment is: **{propagated window type}**<br>- The heater system of this apartment is: **{propagated heating system}**<br>- The energy-efficient lighting of this apartment accounts: **{propagated lighting system}**<br>Treat the background as a WEAK PRIOR that must NOT override clear visual evidence.<br>Output must be a valid JSON object in this format:<br>{{<br>  "epc_category": "one of the following: 'A', 'B', 'C', 'D', 'E', 'F', 'G'"<br>}}<br>Do not include any other text, explanation, or formatting. The entire response must be a valid JSON object. |

Figure 10. Stage-specific prompts for MMCoT